# $AgInSe_2$ Nanorods: A semiconducting material for saturable absorber

Hendry I. Elim and Wei Ji*

Department of Physics, National University of Singapore

2 Science Drive 3, Singapore 117542

Meng-Tack Ng and Jagadese J. Vittal*

Department of Chemistry, National University of Singapore

3 Science Drive 3, Singapore 117543

**Abstract**

$AgInSe_2$ nanorods (NRs) with diameter of 15 nm have been investigated for their nonlinear optical responses by using Z-scan and transient absorption techniques with femtosecond laser pulses of photon energy greater than the bandgap. At excitation irradiance of 20 GW/cm$^2$, $AgInSe_2$ NRs reveal saturation in the nonlinear absorption and optical Kerr nonlinearity with a recovery time determined to be a few ten picoseconds. Such large saturable absorption and Kerr nonlinearity exhibit a third-order susceptibility of 1.2 x $10^{-8}$ esu and a figure of merit of 0.04 esu.cm.s$^{-1}$, making $AgInSe_2$ NRs a promising candidate for saturable absorption devices.

*Electronic mail: phyjiwei@nus.edu.sg or chmjjv@nus.edu.sg

In the past several years, there have been increasing research efforts in the field of nanoscale science and technology toward applications in the area of magnetism, optics and conductivity. With ultrafast laser excitation, semiconducting or metallic nanomaterials have been widely investigated for various optoelectronic applications.[1-4] As an example, semiconducting nanocrystals (NCs) such as CdSe[5], InP[6], or PbS[7] are well studied for their nonlinear optical (NLO) responses to laser pulses of duration from hundred femtoseconds to ten picoseconds. In the spectral regime where the photon energy ($\hbar\omega$) is less than the bandgap ($E_g$), two photon absorption and two-photon-absorption-associated processes are dominant mechanisms. When $\hbar\omega > E_g$, however, saturable absorption due to band filling mechanism plays an important role. Metallic gold particles or nanorods (NRs) have been found to possess strong nonlinear absorption, nonlinear scattering, and considerable local-field enhancement occurring at the surface plasmon resonance (SPR).[8-10] Moreover, semiconducting bundle carbon nanotubes also exhibit ultrafast yet giant NLO effects at the resonant bands, showing great potential for all-optical switching,[11,12] or saturable absorbers[13,14]. These studies are indicative of the superiority of one-dimensional over zero-dimensional nanomaterials. Other interesting properties of one-dimensional nanomaterials are their excellent emission quantum yield and photostability. Enhancement by over one order of magnitude in CdSe NR photoluminescence quantum yield and stabilization against surface oxidation have been obtained by epitaxial overgrowth of an inorganic shell of a higher band gap material.[15,16] In this Letter, we report the optical studies of semiconducting $AgInSe_2$ NRs. Our femtosecond Z-scan and time-resolved transient absorption measurements are indicative of their advantage over semiconducting nanocrystals within the context of all-optical switching.

Colloidal NCs may be prepared by a so-called injection method. This method has led to successful synthesis of a variety of high-quality binary NCs.[17] However, the injection-based synthetic method usually involves elaborate preparation of air-sensitive organometallic complexes at relatively high temperature. Furthermore, such injection method has not been extended to synthesize ternary NRs. Here we present the one-pot synthesis of $AgInSe_2$ NRs, which is divided into three parts as shown in a scheme below:

4PhC{O}Se-Na+ + $InCl_3$ --> [NaIn(SeC{O}Ph)$_4$] + 3NaCl

[NaIn(SeC{O}Ph)$_4$] + (Ph3P)$_2$AgNO$_3$ --> [(Ph3P)$_2$AgIn(SeC{O}Ph)$_4$] + NaNO$_3$

[(Ph3P)$_2$AgIn(SeC{O}Ph)$_4$] ------------------------------> $AgInSe_2$ NRs
thermolysis in OA/DT

The first part is the synthesis of [NaIn(SeC{O}Ph)$_4$]. Briefly, the MeCN in $Na^+$PhC{O}$Se^-$ (9.05 mmol) solution[18] was removed to dryness under vacuum. 30 mL of degassed $H_2O$ was then added to the crude sodium monoselenocarboxylate and the insoluble ppt was filtered off. $InCl_3$ (0.50 g, 2.26 mmol) dissolved in degassed $H_2O$ (10 mL) was added dropwise to the yellow colour filtrate and a pale yellow ppt formed immediately. The solution was stirred for 1.5 hours at room temperature and the resultant white colour ppt was filtered off. The product was washed with plenty of $H_2O$, then dried under vacuum and stored at 5 °C for further use. [Yield: 1.55 g (78 %). Elemental Anal: Calcd. for $NaInSe_4C_{28}H_{20}O_4$ (mol wt 874.11): C, 38.47; H, 2.31 %. Found C, 37.69; H, 2.47 %. $^1$H NMR ($d_6$-Acetone) $\delta_H$: 8.01 – 8.03 (8H, d, $J$ = 6 Hz, *ortho*-proton), 7.38 – 7.42 (8H, t, $J$ = 6 Hz, *meta*-proton), 7.49 – 7.53 (4H, t, $J$ = 6 Hz *para*-proton). $\delta_c$ ($d_6$-Acetone): For selenobenzoate ligand: 127.12 ($C_{2/6}$ or $C_{3/5}$), 128.61 ($C_{2/6}$ or $C_{3/5}$), 131.05 ($C_4$), 142.94 ($C_1$), 201.50 (<u>C</u>OSe). ). ESI-MS (m/z)(Acetone): 898.5 ([NaIn(SeC{O}Ph)$_4$] + $Na^+$, 100%), 852.8 ([In(SeC{O}Ph)$_4$]$^-$, 100 %) 185.3 (PhC{O}$Se^-$, 80%)]. The second part is to synthesize the silver indium selenocarboxylate, [($PPh_3$)$_2$AgIn(SeC{O}Ph)$_4$]. It is interesting to note that this compound can only be synthesized in good yield (65%) from the metathesis reaction between [NaIn(SeC{O}Ph)$_4$] and [($PPh_3$)$_2$Ag($NO_3$)],[19] which is different from the corresponding thio-analogues.[20] Briefly, to [NaIn(SeC{O}Ph)$_4$] (0.64 g, of 0.73 mmol) in 5 mL of Acetone, 0.51 g (0.73 mmol) of ($PPh_3$)$_2$Ag($NO_3$) in 5 mL of $CH_2Cl_2$ was added dropwise under ambient conditions. A brownish grey ppt was formed immediately upon stirring. The solution was stirred for 0.5 hour, and the ppt was filtered off, and washed with MeOH, $H_2O$ and $Et_2O$. The product was then dried under vacuum and stored at 5 °C. [Yield: 0.71 g (65 %). Elemental Anal: Calcd. for $AgInSe_4P_2C_{64}H_{50}O_4.CH_2Cl_2$ (mol wt 1568.51): C, 49.77; H, 3.34; P, 3.95 %. Found C, 50.28; H, 3.09; P, 3.89 %. $^1$H NMR ($d_6$-Acetone) $\delta_H$: 7.72 – 7.74 (8H, d, $J$ = 6 Hz, *ortho*-proton), 7.08 – 7.13 (8H, t, $J$ = 15 Hz, *meta*-proton), 7.24 – 7.42 (34H, m, *para*-proton and $PPh_3$). $\delta_c$ ($d_6$-Acetone): For selenobenzoate ligand: 127.12 ($C_{2/6}$ or $C_{3/5}$), 128.61 ($C_{2/6}$ or $C_{3/5}$), 131.05 ($C_4$), 142.94 ($C_1$), 201.50 (<u>C</u>OSe). For $PPh_3$: 128.15 ($C_3$), 129.30 ($C_4$), 133.14 ($C_1$), 133.95 ($C_2$). $\delta_p$ ($d_6$-Acetone): 5.49. ESI-MS (m/z)($CH_2Cl_2$): 631.0 (($PPh_3$)$_2Ag^+$, 100%), 852.8 ([In(SeC{O}Ph)$_4$]$^-$, 100 %) 185.3 (PhC{O}$Se^-$, 80%). TG wt. loss for $CH_2Cl_2$: expected, 5.4 %; found 2.3 %.]. Finally, the compound [($PPh_3$)$_2$AgIn(SeC{O}Ph)$_4$] (0.03 mmol, 50 mg) was added to a flask containing both OA (1.10 mL, 3.37 mmol) and DT (0.81 mL, 3.37 mmol). The precursor dissolved immediately and formed a black solution. This solution was then degassed for 15 min in a vacuum and heated to 185 °C in an oil bath for 17 h in an argon atmosphere. At the end of the reaction, a black precipitate was found at the bottom of the flask. A small amount of toluene and a large excess of EtOH were added to the reaction solution, and $AgInSe_2$ NRs were separated from the reaction solution by centrifugation. The $AgInSe_2$ NRs are freely dispersed in toluene. The synthesis of the $AgInSe_2$ NRs has been briefly reported recently.[21] In Fig. 1(a), the high resolution transmission electron microscopic (HRTEM) image clearly shows the monodispersity of the $AgInSe_2$ NRs and Fig. 1(b) shows their aspect ratio distribution with the average diameter, length and aspect ratio being (14.5 ± 1.8) nm, (50.3 ± 5.0) nm, and (3.5 ± 0.6), respectively.

As displayed in Fig. 1(c), the absorption spectrum of the $AgInSe_2$ NRs dissolved in toluene was measured with a UV-visible spectrophotometer (UV-1700 Shimadzu). It is evident that there is a long tail in the absorption spectrum. This long tail represents that the $AgInSe_2$ NRs are crystalline in nature, which is different from their bulk counterpart.[22] The bandgap of $AgInSe_2$ NRs can be approximated by density functional theory under local density approximation (LDA)[23], $\Delta E_g = E_{g,\mathrm{NRs}} - E_{g,\mathrm{bulk}} \approx 1.95/(d)^{1.24}$, where $E_{g,\mathrm{NRs}}$ is the bandgap of NRs, $E_{g,\mathrm{bulk}}$ (=1.25 eV) is the bandgap of bulk $AgInSe_2$[22] and $d$ is the diameter of the NRs. With $d$ = 14.5 nm, we find the $E_{g,\mathrm{NRs}}$ is located at 1.32 eV.

As displayed in Fig. 2(a), the open-aperture Z-scan data indicate that the nonlinear absorption coefficient in the $AgInSe_2$ NRs is negative due to saturable absorption or optical induced transparency in the $AgInSe_2$ NRs. The similar observation has been also obtained from gold NRs[10] and carbon nanotubes[24]. At laser irradiances higher than 20 GW/cm$^2$, the $AgInSe_2$ NRs reveal obviously saturation in the nonlinear absorption. In this saturation regime, most of the carrier states are filled since the photon energy (=1.59 eV) is larger than the bandgap, and thus the absorption is completely quenched. The similar observation has been carried out in isolated semiconducting single-walled carbon nanotubes (SWNTs) by Ostojic *et al.*[24] Such saturation in the isolated SWNTs was observed in a low fluence of 1 mJ/cm$^2$ (~ 7 GW/cm$^2$) due to the resonance condition. By fitting the experimental data using a hyperbolic saturable absorption function, $\alpha_2^{\mathrm{NRs}} = (\alpha_2^0)_{\mathrm{NRs}}/(1+I/I_s)$, where $\alpha_2^{\mathrm{NRs}}$ is the nonlinear absorption coefficient of the NRs assumed as $\alpha_2^{\mathrm{NRs}} \approx \alpha_2/(f^4 V_{\mathrm{f}})$, where $f$ is the local field factor, $V_{\mathrm{f}}$ the volume fraction of NRs relative to toluene, $(\alpha_2^0)_{\mathrm{NRs}}$ the small-signal absorption coefficient of the NRs, $I$ the measured irradiance, and $I_s$ the saturation irradiance, we determine $\alpha_2$, $(\alpha_2^0)_{\mathrm{NRs}}$ and $I_s$ for the $AgInSe_2$ NRs to be -0.29 cm/GW (or $\mathrm{Im}(\chi^{(3)}) = -2.2 \times 10^{-12}$ esu), -1500 cm/GW (or $\mathrm{Im}(\chi^{(3)})_{\mathrm{NRs}} = -5.4 \times 10^{-9}$ esu) and 20.1 GW/cm$^2$, respectively. This saturation intensity is larger than that for isolated SWNTs.[24] Moreover, such large saturable absorption of the $AgInSe_2$ NRs may be attributed to the intrinsic properties of $AgInSe_2$ NRs. By comparison, a value of $4 \times 10^{-13}$ esu at 820 nm was observed for $|\chi^{(3)}|$ in SWNTs solution at a concentration of 0.33 mg/mL.[25]

Similarly, the saturation in the optical Kerr nonlinearity has also been observed in the $AgInSe_2$ NRs. Figure 2(b) displays examples of the closed-aperture Z-scans of the $AgInSe_2$ NRs carried out with three different excitation irradiances. By fitting the experimental data with the following formula, $n_2^{\mathrm{NRs}} = (n_2^0)_{\mathrm{NRs}}/(1 + I/I_S)$, where $n_2^{\mathrm{NRs}}$ is the nonlinear refractive index of NRs assumed as $n_2^{\mathrm{NRs}} \approx (n_2-(1-V_{\mathrm{f}})n_2^{sol})/(f^4 V_{\mathrm{f}})$, where $n_2^{sol}$ is the nonlinear refractive index of toluene, and $(n_2^0)_{\mathrm{NRs}}$ the small-signal nonlinear refraction coefficient, one finds that $n_2$ is $6.8 \times 10^{-5}$ cm$^2$/GW (or $\mathrm{Re}(\chi^{(3)}) = 3.9 \times 10^{-12}$ esu), and $(n_2^0)_{NRs}$ and $I_s$ values displayed in Fig. 2(c) are 0.35 $cm^2$/GW (or $Re(\chi^{(3)})_{NRs}$ = 1.1 x $10^{-8}$ esu) and 18.0 GW/$cm^2$, respectively. The saturation intensity of the nonlinear refractive index of the $AgInSe_2$ NRs is slightly smaller than that for nonlinear absorption coefficient. By comparison, a value of ~$10^{-9}$ esu at ~1300 nm was observed for $Re(\chi^{(3)})$ in 100 nm-thick layer of SWNTs deposited on a glass substrate.[26] Our smaller value of $Re(\chi^{(3)})$ is due to small volume fraction of $AgInSe_2$ NRs in toluene. In addition, there are few reports of saturable Kerr nonlinearity. The first one was in single-crystal para-toluane sulfonate (PTS). Torruellas *et al.* reported that the values of $n_2^0$ and $I_s$ of single-crystal PTS measured at 730 nm are -4.9 $cm^2$/GW and 1.9 GW/$cm^2$, respectively.[27] Such large saturable Kerr nonlinearity is due to a strong two-photon band accessible in the bandgap of PTS.

As an example displayed in Fig. 3(a), typical transient transmission measurements show a positive change in the transmission which results from photoinduced bleaching. At a very high irradiance of ~20 GW/$cm^2$, the $AgInSe_2$ NRs reveal saturation. This result is consistent with the previously-discussed Z-scan results. The pump-probe data fitted with a two-exponential component model reveal that the photo-excited electron dynamics in the $AgInSe_2$ NRs measured at excitation irradiances smaller than 20 GW/$cm^2$ is dominated by two relaxation processes: the fast (3 ps) and the slow (20 ps) decay components. The fast dynamics is commonly attributed to trapping from shallow trap states to deep trap states, whereas the slow dynamics is assigned to electron-hole combination for deep trap states. The deep states in the bandgap may be related to excess Ag atoms at the interface of $AgInSe_2$. Such dynamics mechanism is similar with PbS nanocrystals.[28] Further studies far beyond the scope of this investigation are required to unambiguously identify the nature of the relaxation in the $AgInSe_2$ NRs. In addition, it should be pointed out that the relaxation dynamics of the $AgInSe_2$ NRs are dependent on the pump irradiance. At pump irradiances larger than 20 $GW/cm^2$, the slow decay component is prolonged (> 20ps) due to saturation process in which more carrier states are filled and thus the absorption is completely quenched. Figure 3(b) shows the pump irradiance dependence of the maximum value of $\Delta T / T$ in the $AgInSe_2$ NRs. This is in agreement with the Z-scan results in Fig. 2(c), showing that the nonlinear responses of the $AgInSe_2$ NRs are dominated by a saturation process. Table 1 tabulates the NLO parameters of $AgInSe_2$ NRs and other nanomaterials. The figure of merit (FOM), $|\chi^{(3)}|/(\alpha_0\tau)$ of the $AgInSe_2$ NRs is 0.04 $esu.cm.s^{-1}$, about one order larger than metallic Au NRs[10] and few times larger than CdSe NCs[5] or PbS NCs[7]. However, its FOM is at least one order less than that of SWNTs[11,25] and several orders smaller than the InGaAs/AlAs/AlAsSb-coupled quantum wires (QWs)[29] due to a relatively slower response of the $AgInSe_2$ NRs.

In conclusion, we have reported the synthesis of $AgInSe_2$ NRs which are ~50 nm in length and ~15 nm in diameter. We have also observed their large, ultrafast saturable absorption and optical Kerr nonlinearity. The third-order nonlinear optical susceptibility, $\left|\chi^{(3)}\right|_{\mathrm{NRs}} = \sqrt{\left(\mathrm{Im}\left(\chi^{(3)}\right)_{\mathrm{NRs}}\right)^2 + \left(\mathrm{Re}\left(\chi^{(3)}\right)_{\mathrm{NRs}}\right)^2}$ of the $AgInSe_2$ NRs is determined to be 1.2 x $10^{-8}$ esu at 780 nm wavelength. Our measurements reveal that the recovery time of photo-excited electrons in the $AgInSe_2$ NRs is about 20 ps. It should be pointed out that the observed saturable absorption makes $AgInSe_2$ NRs promising for ultrafast saturable absorber.

**Table 1.** Nonlinear Parameters and Figure of Merit (FOM).

| Materials | $\alpha_0$ (cm$^{-1}$) | $\alpha_2$ (cm/GW) | $\alpha_2^{NRs/QDs}$ (cm/GW) | $n_2^{NRs/QDs}$ (cm$^2$/GW) | Im( $\chi^{(3)}$ ) (esu) | FOM = $\lvert\chi^{(3)}\rvert/(\alpha_0\tau)$ (esu.cm.s$^{-1}$) |
|---|---|---|---|---|---|---|
| $AgInSe_2$ NRs | 6.1 | -0.29 | -1500 | 0.35 | -2.2 x 10$^{-12}$ | 0.04 |
| CdSe [a)] | | | | | -6.0 x 10$^{-8}$ | |
| PbS [b)] | 1.5 | -0.75 | | | -1.1 x 10$^{-12}$ | 0.002 |
| Au NRs [c)] | 3 | -1.5 | | | -1.2 x 10$^{-12}$ | 0.004 |
| SWNTs [d)] | 65000 | | | | -0.8 x 10$^{-7}$ | 2.4 |
| SWNTs [e)] | 1.86 | | | | -4.0 x 10$^{-13}$ | 0.53 |
| InGaAs/AlAs /AlAsSb[f)] | | | | | | 174 |

a) Ref. 5, b) Ref. 7, c) Ref. 10, d) Ref. 11, e) Ref. 25, and f) Ref. 29.

**Figure Captions:**

**Figure 1.** (**a**) TEM images of the $AgInSe_2$ NRs. (**b**) Counts of the $AgInSe_2$ NRs versus aspect ratio. (**c**) UV-Vis absorption spectrum of the $AgInSe_2$ NRs dissolved in toluene with a volume fraction ($V_f$) of 3.9 x $10^{-4}$.

**Figure 2.** (**a**) Open- and (**b**) closed-aperture Z-scans of 1-mm-thick solution of the $AgInSe_2$ NRs measured with 780-nm, 200-fs laser pulses at 1-kHz repetition rate. The laser irradiance used is in the range from 5 GW/cm$^2$ to 47 GW/cm$^2$. The solid lines are the best-fit curves calculated by using the Z-scan theory. The closed-aperture Z-scan curves in (**b**) are shifted vertically for clear presentation. (**c**) Irradiance dependence of the nonlinear absorption coefficient ($\alpha_2^{NRs}$) and nonlinear refractive index ($n_2^{NRs}$) for the $AgInSe_2$ NR solution.

**Figure 3.** (**a**) Transient transmission measurements on the $AgInSe_2$ NR solution conducted with 780-nm, 200-fs laser pulses at 1-kHz repetition rate. The solid lines are the best fits that give decay times $\tau_1$ = ~3 ps and $\tau_2$ = ~20 ps for input irradiances ($I$) less than of 20.0 GW/cm$^2$. (**b**) Maximum transmission change versus pump irradiance.

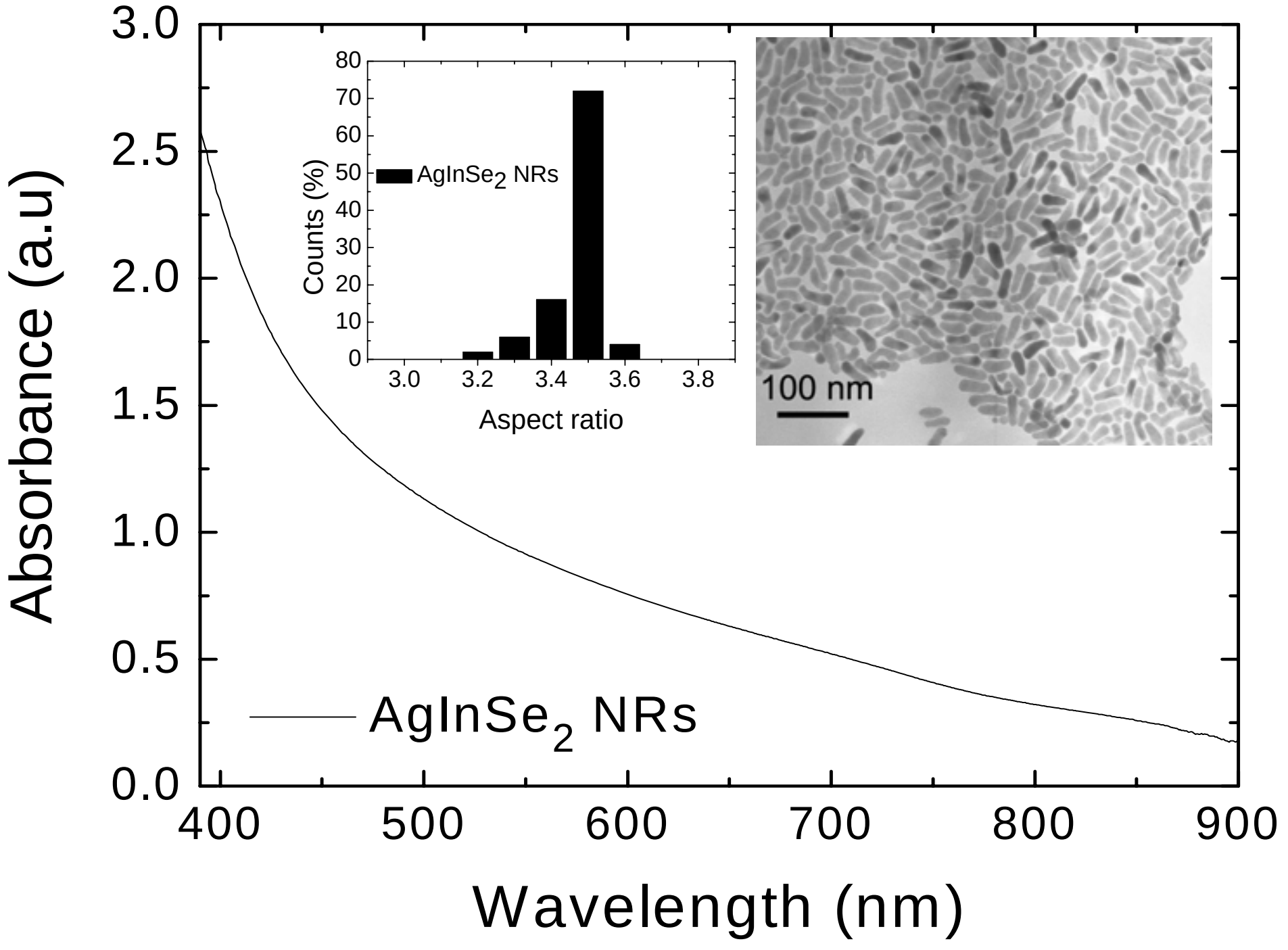

**Fig. 1.**

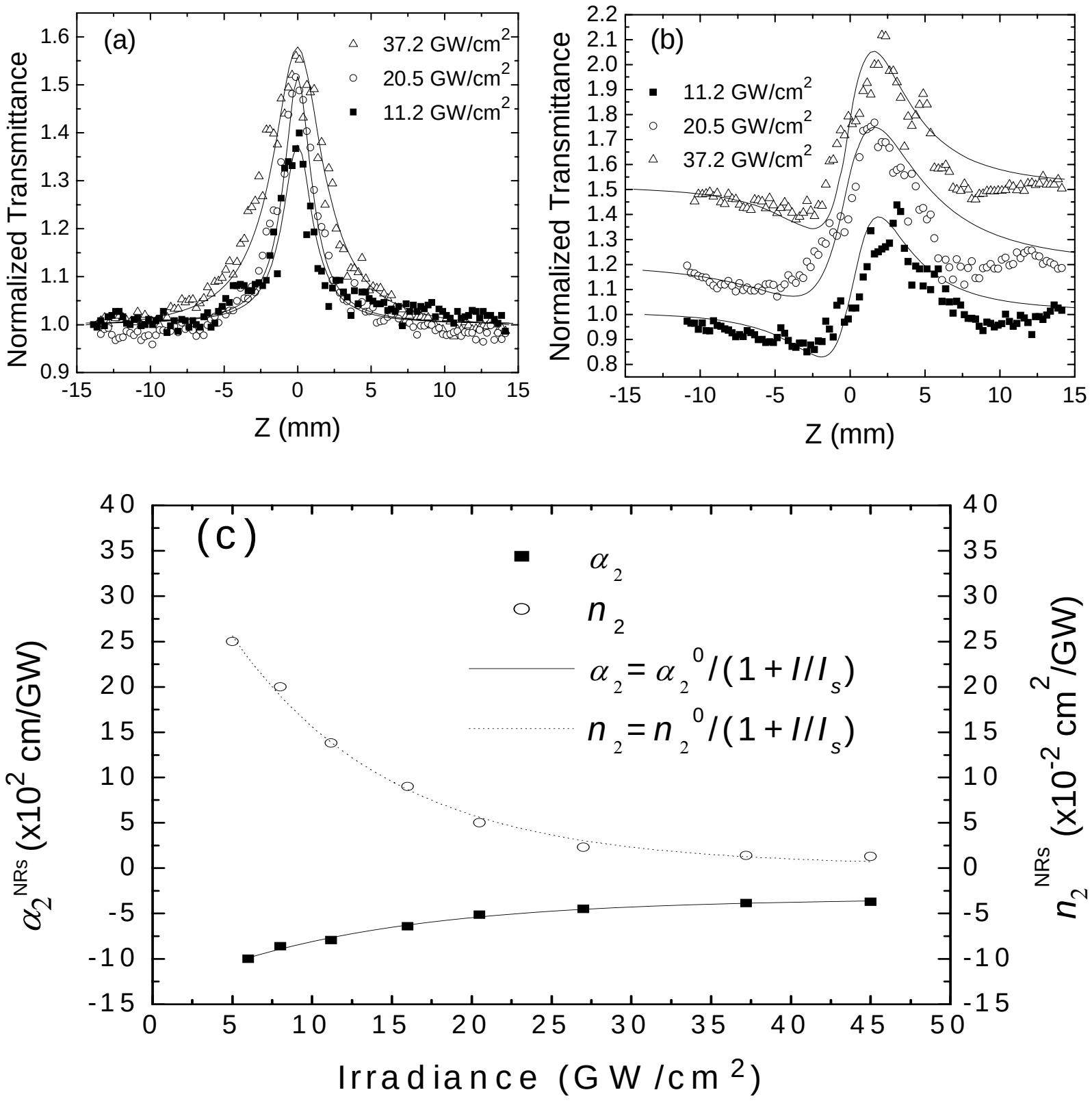

**Fig. 2.**

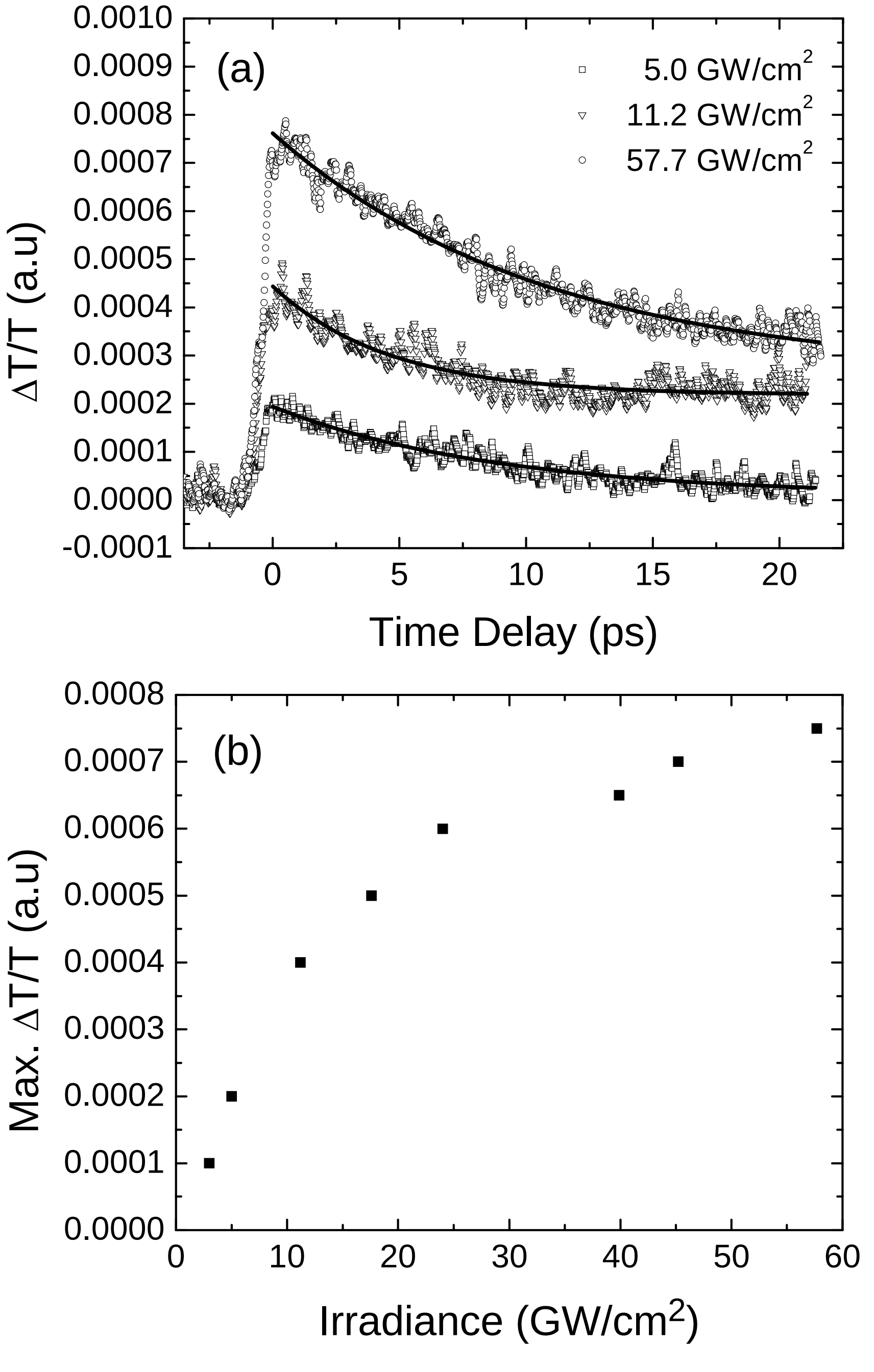

**Fig. 3.**